\pdfoutput=1

\documentclass[11pt]{article}
\usepackage[margin=1in]{geometry}
\usepackage{graphicx}
\usepackage{amsmath,amssymb}
\usepackage{booktabs}
\usepackage{enumitem}
\usepackage{microtype}
\usepackage[T1]{fontenc}
\usepackage[utf8]{inputenc}
\usepackage{url}
\usepackage{float}
\usepackage{abstract}

\DeclareUnicodeCharacter{200B}{}      % ZWSP 제거
\DeclareUnicodeCharacter{200C}{}      % ZWNJ
\DeclareUnicodeCharacter{200D}{}      % ZWJ
\DeclareUnicodeCharacter{2060}{}      % WORD JOINER
\DeclareUnicodeCharacter{FEFF}{}      % BOM
\DeclareUnicodeCharacter{2460}{(1)}   % ①
\DeclareUnicodeCharacter{2461}{(2)}   % ②
\DeclareUnicodeCharacter{2462}{(3)}   % ③

\usepackage[hidelinks]{hyperref}
\title{Persode: Personalized Visual Journaling with Episodic Memory-Aware AI Agent}

\makeatletter
\renewcommand*{\thefootnote}{\fnsymbol{footnote}}
\makeatother

\author{%
\begin{minipage}{0.94\textwidth}\centering
\textbf{Seokho Jin\textsuperscript{1}}, \textbf{Manseo Kim\textsuperscript{1}}, \textbf{Sungho Byun\textsuperscript{1}},\\
\textbf{Hansol Kim\textsuperscript{1}}, \textbf{Jungmin Lee\textsuperscript{1}}, \textbf{Sujeong Baek\textsuperscript{2}},\\
\textbf{Semi Kim\textsuperscript{2}}, \textbf{Sanghum Park\textsuperscript{2}}, \textbf{Sung Park\textsuperscript{3}}\thanks{Corresponding author: \texttt{sjp@taejae.ac.kr}}\\[0.5em]
{\small
\textsuperscript{1}\,Sangmyung University, Republic of Korea\\
\textsuperscript{2}\,Peaknic Co., Republic of Korea\\
\textsuperscript{3}\,Taejae University, Seoul, Republic of Korea
}
\end{minipage}
}

\date{} % 날짜 숨김

\begin{document}
\maketitle

% 이후 각주 표식을 숫자로 되돌려 수식 번호 등에 기호가 섞이지 않게
\renewcommand{\thefootnote}{\arabic{footnote}}
\setcounter{footnote}{0}

%==================================================
% NOTE (figures directory & filenames for arXiv):
%   figures/fig1.png   figures/fig2.png   figures/fig3.png   figures/fig4.png
% Allowed filename chars: a-z A-Z 0-9 _ + - . , = (no spaces / non-ASCII, case-sensitive)

\begin{abstract}
Reflective journaling often lacks personalization and fails to engage Generation Alpha and Z, who prefer visually immersive and fast-paced interactions over traditional text-heavy methods. Visual storytelling enhances emotional recall and offers an engaging way to process personal experiences. Designed with these digital-native generations in mind, this paper introduces Persode, a journaling system that integrates personalized onboarding, memory-aware conversational agents, and automated visual storytelling. Persode captures user demographics and stylistic preferences through a tailored onboarding process, ensuring outputs resonate with individual identities. Using a Retrieval-Augmented Generation (RAG) framework, it prioritizes emotionally significant memories to provide meaningful, context-rich interactions. Additionally, Persode dynamically transforms user experiences into visually engaging narratives by generating prompts for advanced text-to-image models, adapting characters, backgrounds, and styles to user preferences. By addressing the need for personalization, visual engagement, and responsiveness, Persode bridges the gap between traditional journaling and the evolving preferences of Gen Alpha and Z.
\end{abstract}

% % ---- CCS CONCEPTS & KEYWORDS  ----
% \noindent\textbf{CCS CONCEPTS} Human-centered computing $\to$ Interaction design $\to$ Interaction design process and methods; Computing methodologies $\to$ Artificial intelligence $\to$ Natural language processing $\to$ Knowledge representation and reasoning; Information systems $\to$ Information retrieval $\to$ Personalization

% \medskip
% \noindent\textbf{Additional Keywords and Phrases:} Personalized Journaling, Episodic Memory Integration, Emotion-Driven AI, AI-Augmented Reflection, Adaptive Storytelling, Visual and Narrative Synergy, Contextual Memory Retention, User-Centered AI Design, Retrieval-Augmented Generation (RAG), Emotion-Aware Interactions

\section{Introduction}
Reflective journaling has long been recognized as a powerful tool for enhancing self-awareness, emotional regulation, and mental well-being, serving as a cornerstone in mental health practices across diverse age groups [3]. However, digital-native generations such as Generation Alpha and Z prefer visually immersive, socially connected, and fast-paced interactions, making traditional, text-heavy journaling less engaging [1, 7].

Although highly fluent in digital media, Generation Alpha and Z show diminished interest in traditional reflective journaling, preferring platforms that provide immediate feedback, visual engagement, and social validation. Research shows that while Generation Alpha and Z frequently express personal thoughts through brief social media posts or multimedia snippets, they seldom engage in structured, long-form journaling, perceiving it as less adaptable to their fast-paced routines [16]. The absence of instant feedback and delayed gratification further discourage Generation Alpha and Z from engaging with traditional journaling practices. With a preference for visually engaging content, Generation Alpha and Z often find traditional text-based journaling unappealing [1].

To address these evolving preferences, we propose Persode -- a system that integrates episodic memory-aware conversational agents with AI-driven visual storytelling, delivering a personalized and engaging journaling experience. By blending personalized onboarding, memory-optimized dialogues, and dynamic visual storytelling, Persode bridges the gap between traditional journaling practices and the expectations of digital-native generations. It aims to create an engaging, reflective space where users can explore their memories and emotions through a rich combination of narrative and visuals, tailored to their individual preferences.

\section{RELATED WORKS}
AI-driven journaling systems have shown promise in enhancing self-reflection, emotional expression, and mental well-being, gaining attention from researchers and practitioners [5, 6, 10, 19]. Large language models (LLMs) contribute to reflective prompts and consistent journaling habits but face challenges in addressing nuanced emotional and situational contexts [4, 5]. Despite offering personalized recommendations, many tools fail to deeply resonate with users, often struggling to capture the emotional depth required for meaningful engagement [5, 6, 11]. Moreover, over-reliance on AI-generated text can diminish user authenticity and engagement, highlighting the critical challenge of balancing AI assistance with meaningful user autonomy [4, 6]. 

Generations Alpha and Z prefer brief, visually dynamic content over structured journaling, reflecting their fast-paced digital habits [1]. Their reliance on algorithm-driven recommendations and instant gratification further reduces the appeal of traditional reflective journaling tools [14]. Research also highlights their preference for visual and interactive journaling experiences, consistent with broader trends in digital consumption habits [7]. These preferences emphasize the need for systems that integrate adaptability, visual engagement, and responsiveness to evolving user expectations.

Existing journaling platforms, such as Draw My Day [18], Reflectr [2], and Reflectly [15], incorporate interesting features but often fall short in achieving emotional depth or providing integrated multimedia journaling experiences. Their reliance on rigid templates limits their ability to foster meaningful and personalized user interactions. This suggests a gap in current systems, where deeper emotional resonance and dynamic, user-centered designs are needed.

Our work addresses these gaps by examining journaling as an emotionally resonant and identity-building activity. We emphasize the significance of emotional continuity, contextual adaptability, and self-identity reinforcement, drawing on psychological research regarding emotional arousal, contextual relevance, and the reinforcement of significant memories [17]. Rather than relying solely on prompt generation or predefined templates, we explore how systems can dynamically adapt to users' emotional states, offering cohesive narrative experiences that integrate text and visuals to support personal growth and meaningful reflection over time. The following section introduces the core design concept and system architecture, demonstrating how these elements address the limitations identified in prior research and relevant services.

\section{System Design overview}

\subsection{Core Design Concepts}
The name ``Persode,'' a fusion of ``personalized'' and ``episode,'' reflects its focus on delivering personalized and emotionally resonant journaling experiences. Users begin their journey by customizing their persona’s appearance -- such as hair color or glasses -- and adjusting the chatbot’s personality to match their preferences. This onboarding process prepares users for meaningful interactions tailored to their unique identity.

As users engage with the chatbot, the conversational agents, guided by an episodic memory framework, encourage reflective dialogues. These dialogues dynamically adapt to the user’s emotional state and context, extracting key elements (e.g., events, people, objects, places, and emotions) that are central to their experiences. For example, recalling a family gathering enables the system to capture details like ``family,'' ``dinner table,'' and ``joyful atmosphere,'' transforming them into a cohesive narrative enriched with matching text and visuals.

By leveraging these episodic memories, the system retrieves similar past experiences to enhance its responses with empathetic feedback, helping users reflect more deeply. These insights are seamlessly transformed into reflective diary entries and personalized illustrations, aligning with the user’s emotions and context. By presenting both text and visuals in a cohesive journal format, Persode fosters deep reflection, helping users reconnect with their memories and emotions.

\subsection{Episodic Memory-Aware Conversational Agent}
The absence of long-term memory mechanisms in AI systems limits their ability to retain and leverage historical interactions, reducing their effectiveness in tasks requiring sustained engagement and personalization [20]. Persode employs a Memory-Strength Scoring Mechanism inspired by Ebbinghaus’ Forgetting Curve [9]. The system references short-term memories within six days, reflecting the steepest retention drop of approximately 75\% observed in his findings. Beyond this, long-term memories are evaluated based on emotional intensity, recall frequency, and contextual relevance, enabling the retention of significant events while optimizing computational efficiency. Our approach incorporates memory decay principles to emulate natural forgetting while drawing from Craik and Lockhart’s Levels of Processing Theory [21], which suggests that deeper and more meaningful processing such as emotionally significant or contextually relevant experiences leads to stronger and longer-lasting retention. By emphasizing emotionally salient and contextually rich memories, Persode prioritizes information that holds personal meaning for users, enhancing engagement and relevance. 

To bridge short-term and long-term memory, Persode employs a Retrieval-Augmented Generation (RAG) mechanism [8], which enhances response accuracy by retrieving relevant context from stored information. This allows the system to reference recent events for immediate situational awareness while integrating emotionally significant long-term memories. By dynamically aligning text and visuals with the user’s emotional context, Persode fosters a cohesive journaling experience that supports sustained engagement and personalization.

Inspired by models like MemoryBank [20] and LUFY[17], this mechanism integrates relevant fragments of past interactions in real-time (see Figure 1.A), enabling accurate and emotionally resonant responses. To further personalize user experiences, the system also incorporates onboarding preferences specified by the user. During onboarding, users customize chatbot response styles by selecting options such as empathetic tones, detailed answers, or direct expressions. Based on these preferences, the system automatically generates and combines tailored prompts to guide response generation. For instance, selecting highly emotional and empathetic settings adjusts the chatbot’s tone and phrasing to align with these choices (see Figure 1.B).

By integrating memory scoring with RAG, the system enables adaptive flows that dynamically retrieve and incorporate emotionally significant memories. These flows enhance personalization by aligning responses with user behavior and emotional context, while ensuring contextual relevance through effective retrieval strategies. This balanced approach optimizes memory retention and computational efficiency.

% ---- Figure 1 (inline after first mention) ----
\begin{figure}[H]
  \centering
  \includegraphics[width=0.95\linewidth]{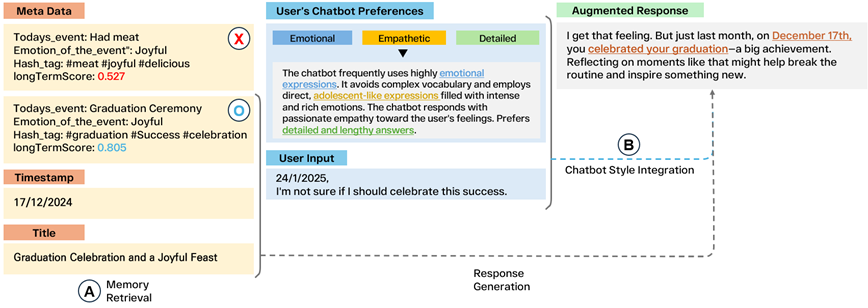}
  \caption{Integration of chatbot preferences and memory retrieval for personalized responses.}
  \label{fig:integration}
\end{figure}

\subsection{Dual-Template Episodic Memory Generation Framework}
Our system is designed to enhance journaling by providing personalized text-based diary entries and visuals, fostering a dual approach to deeper emotional engagement. By centering on the user’s reflective journey, the system creates meaningful opportunities for users to reconnect with their memories in a more impactful and resonant way.

The text template helps users revisit significant moments through diary entries that distill the emotional essence of their experiences. Rather than serving as mere event logs, these entries provide space for reflection, allowing users to explore the emotional dimensions of their personal milestones or everyday interactions. For example, in Figure 1, students may find new perspectives on their academic progress, while others might use the entries to better understand the emotions behind cherished moments.

The visual template complements this by translating these memories into personalized visuals that align with the user’s emotions and stylistic preferences. These visuals aim to serve as a medium through which users can reconnect with their narratives in a more immersive and creative way. For instance, a visual representation of a serene sunset might evoke tranquility and introspection, offering users a fresh perspective on their experiences. While the impact of this approach may vary among individuals, it holds the potential to provide an engaging and reflective journaling experience.

By aligning outputs with users’ emotional contexts and preferences, the framework seeks to make journaling more accessible and personally meaningful. Although its effectiveness in fostering emotional connection will benefit from further evaluation, the system’s design is rooted in the idea that personalized, emotionally resonant outputs can support self-reflection and creative expression.

\section{Implementation}
Persode's implementation integrates advanced AI-driven components with a scalable backend architecture to deliver a seamless and emotionally engaging journaling experience. The system’s architecture, illustrated in Figure 2, combines episodic memory-aware conversational agents with personalized visual storytelling, ensuring efficient data flow and alignment with user preferences.

% ---- Figure 2 (inline after first mention in Implementation) ----
\begin{figure}[H]
  \centering
  \includegraphics[width=0.95\linewidth]{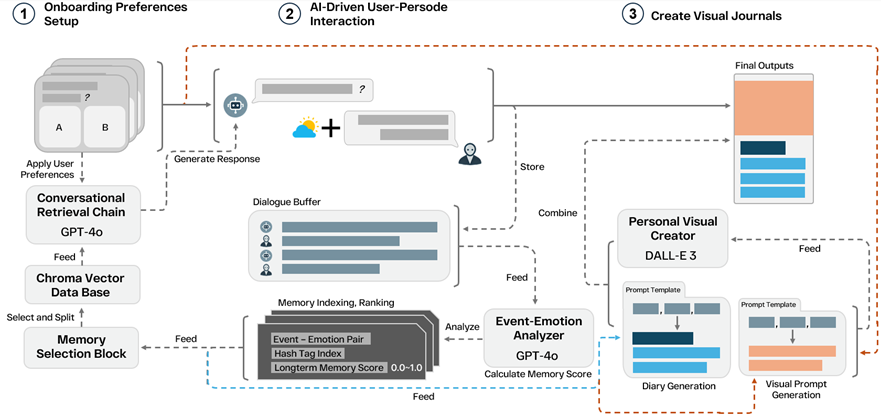}
  \caption{Overview of Persode's system architecture, combining user onboarding, memory-driven conversational agents, and AI-powered visual storytelling for an emotionally engaging journaling experience.}
  \label{fig:arch}
\end{figure}

\subsection{Onboarding User Preferences}
The onboarding process, as shown in Figure 2-① and Figure 3, is designed to personalize every aspect of the journaling experience. Users begin by defining their stylistic and demographic preferences, selecting elements such as glasses, fashion styles (see Figure 3.b), and background aesthetics (see Figure 3.c). These inputs directly inform the text-to-image generation pipeline, ensuring that all visuals align with the user’s unique identity. Additionally, users customize their chatbot’s personality by selecting conversational traits, such as ``empathetic'' or ``friendly'' tones (see Figure 3.d). This customization ensures that the chatbot’s interactions align with the user’s emotional preferences, enhancing the overall experience. All onboarding data is stored in the database and dynamically referenced during conversations and visual output generation. The visual storytelling pipeline ensures that generated images align not only with the user's narrative but also with their unique identity and stylistic preferences, enhancing both emotional connection and visual coherence.

% ---- Figure 3 (inline inside Onboarding) ----
\begin{figure}[H]
  \centering
  \includegraphics[width=0.95\linewidth]{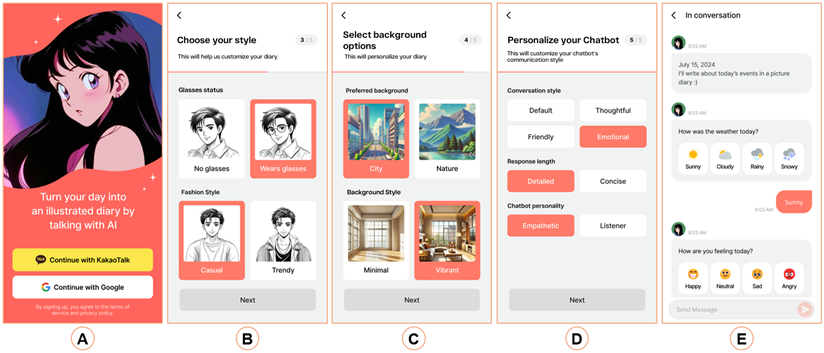}
  \caption{Onboarding process allowing users to personalize stylistic preferences, chatbot personality, and visual settings for a tailored journaling experience.}
  \label{fig:onboarding}
\end{figure}

\subsection{Emotion-Driven Episodic Memory in Conversational AI}
At the heart of Persode’s system lies the episodic memory-aware conversational agent, which dynamically integrates past interactions and emotional contexts into ongoing dialogues (refer to Figure 2-②). User dialogue inputs are stored in a temporary buffer, which is then analyzed by the Event-Emotion Analyzer. This module extracts structured metadata such as event-emotion pairs, timestamps, and hashtags, transforming raw conversational data into actionable insights. Event-emotion pairs are organized using episodic memory, associating significant events with emotional states. Long dialogues are segmented into discrete events based on episodic memory principles, ensuring clarity and emotional context.

To manage memory effectively, the extracted metadata undergo evaluation through the Memory Strength Scoring Mechanism, which prioritizes memories based on emotional intensity, recall frequency, and contextual relevance.

% --- Equation (1): normalized weighted score with optional time decay ---
\begin{equation}
\label{eq:score}
S \;=\; d(\Delta t)\,\cdot\,
\frac{w_E\,E \;+\; w_R\,R \;+\; w_C\,C}{\,w_E \;+\; w_R \;+\; w_C\,},
\end{equation}

\noindent
where $E$ is emotional intensity, $R$ is recall frequency, and $C$ is contextual relevance.
$w_E,w_R,w_C \ge 0$ are tunable weights (not necessarily summing to 1), and
$d(\Delta t)\in(0,1]$ is a time-based decay that monotonically decreases with the elapsed time
$\Delta t$ since the memory was formed (e.g., an exponential decay $d(\Delta t)=e^{-\lambda \Delta t}$).
When $d(\Delta t)=1$ or is absorbed into $R$, Equation~(\ref{eq:score}) reduces to the normalized
weighted average used at retrieval time.

During The Memory Selection Block utilizes these stored fragments to identify and retrieve relevant past experiences dynamically. By referencing these memories, the system augments ongoing dialogues with contextually meaningful insights. For instance, if a user recalls a joyful celebration, the agent may reference a similar prior event, such as a graduation ceremony, to enrich the conversation. This not only deepens emotional resonance but also maintains continuity across interactions.

\subsection{Creating Illustrated Diaries}
The visual generation system complements the conversational agent by transforming textual reflections into personalized diary entries and imagery (refer to Figure 2-③). Textual prompts for diary entries are generated by GPT-4o [13] which analyzes the dialogue content to reflect the user's emotional state and stylistic preferences. These diary entries are concise yet tailored, designed to encapsulate the essence of the user’s experiences. On average, diary generation takes approximately three seconds, ensuring responsiveness without sacrificing quality.

For visual outputs, DALL-E 3 [12] translates event-emotion metadata and onboarding preferences into customized imagery. This process leverages the Few-Shot Template System (refer to Figure 2-③, lower section), which combines user-defined onboarding preferences (e.g., age, preferred fashion styles, and background themes) with metadata extracted from user conversations. The Few-Shot Template System automates the transformation of these inputs into detailed image prompts. For example, when a user reflects on a moment involving a teenage girl with dyed yellow hair who is being scolded by her mother for spending all her allowance, the system generates a visual narrative. Incorporating the emotional context (``sorrowful but reflective'') and stylistic details (casual attire and dyed hair), the output depicts the girl in a scene highlighting her feelings of regret and emotional intensity. This ensures the user's recollection is enriched with a personalized visual that aligns seamlessly with the reflective diary entry. 

\subsection{From Conversation to Visual Diary}
Figure 4 illustrates the end-to-end process of transforming dialogue into a cohesive illustrated diary entry. As users engage in reflective conversations, key events and emotions are extracted and encoded into structured representations (see Figure 4.C). In this example, a user shares their feelings about an upsetting incident involving a car splashing water on them, which ruined their favorite outfit (refer to Figure 4.A, B). The system extracts key emotions like frustration and sadness, along with events such as ``ruined outfit'' and contextual hashtags (\#FavoriteOutfit, \#Upset, \#Laundry).

Through the Few-Shot Template System, this metadata is combined with the user’s onboarding preferences, such as preferred stylistic elements and narrative tone (refer to Figure 4.D). For instance, the diary entry reflects both the emotional weight of the event and the visual representation of laundry-related imagery, featuring a character washing clothes with a contemplative expression (refer to Figure 4.E).

The final output harmonizes the GPT-4o-generated textual narrative with a visually evocative scene created by DALL-E 3, ensuring that the diary entry resonates both emotionally and aesthetically with the user. This process showcases how the system captures the user's unique experiences and styles to craft a personalized and engaging journaling artifact.

% ---- Figure 4 (inline after first paragraph of this subsection) ----
\begin{figure}[H]
  \centering
  \includegraphics[width=0.95\linewidth]{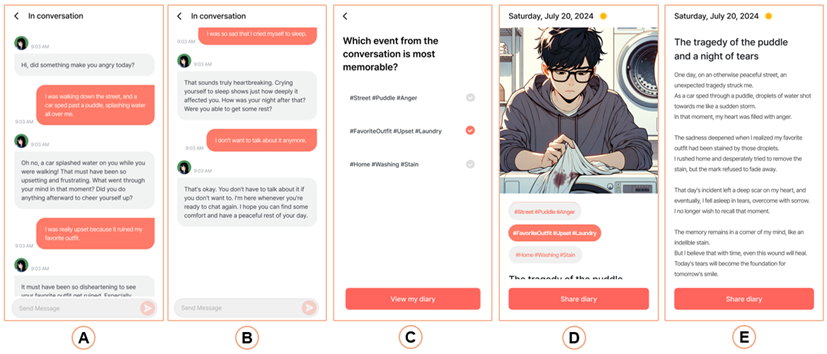}
  \caption{Dialogue and diary generation process, illustrating how user conversations are transformed into structured events, visual prompts, and personalized illustrated diary entries.}
  \label{fig:pipeline}
\end{figure}

\section{CONCLUSION and discussion}
Persode underscores the importance of user-centered design in developing reflective journaling systems. By integrating personalized onboarding, memory-optimized conversational agents, and automated visual storytelling, the system creates a space where users can deeply engage with their personal narratives in ways that feel authentic and meaningful. Features that prioritize user preferences and emotional context address common pain points in journaling tools, such as lack of relevance and engagement.

Moving forward, we plan to conduct extensive user testing and deploy the system in real-world scenarios to gather diverse user feedback and collect valuable data on performance, usability, and overall impact. This phase will inform iterative improvements to better meet user needs and ensure the platform evolves effectively to accommodate diverse requirements.

Additionally, scalability and adaptability in various contexts will be explored to ensure the system's robustness and accessibility across different demographics. Future research will focus on longitudinal studies to understand the long-term effects of Persode on user engagement and satisfaction. By analyzing user interactions and responses, the system will be refined to foster deeper emotional resonance and greater accessibility.

By focusing on accessibility, emotional resonance, and personalized experiences, Persode aspires to empower individuals to reflect, express, and grow through an enriched journaling journey tailored to their unique needs. Through these efforts, the system aims to contribute to the broader field of personalized AI-driven journaling tools, paving the way for more meaningful and impactful interactions.

\end{document}